\documentclass[aps,pre,showpacs,twocolumn,amsmath,amssymb,superscriptaddress,floatfix]{revtex4-2} 
\usepackage{graphicx}
\usepackage{dcolumn}
\usepackage{bm}
\usepackage{color}

\newcommand{\gen}{t}
\newcommand{\G}{\mathcal{G}}
\newcommand{\Ggen}{G_{\text{gen}}}
\newcommand{\mgenav}{\langle m_{\text{gen}}\rangle}
\newcommand{\nremav}{\langle n_{\text{rem}}\rangle}
\newcommand{\kappaav}{\langle \kappa\rangle}
\newcommand{\lambdaav}{\langle \lambda\rangle}
\newcommand{\Df}{D_{\text{f}}}
\newcommand{\df}{d_{\text{f}}}
\newcommand{\dfmin}{d_{\text{f}}^{\text{min}}}
\newcommand{\dfmax}{d_{\text{f}}^{\text{max}}}

\begin{document}

\preprint{APS/123-QED}

\title{Bifractality of fractal scale-free networks}

\author{Jun Yamamoto}
\email{j.yamamoto@se22.qmul.ac.uk} \email{jun.j.yamamoto@gmail.com}
\affiliation{%
Department of Applied Physics, Hokkaido University, Sapporo 060-8628, Japan}
\affiliation{%
School of Mathematical Sciences, Queen Mary University of London, London E1 4NS, United Kingdom}
\author{Kousuke Yakubo}%
\email{yakubo@eng.hokudai.ac.jp}
\affiliation{%
Department of Applied Physics, Hokkaido University, Sapporo 060-8628, Japan}
\date{\today}

\begin{abstract}
The presence of large-scale real-world networks with various architectures has motivated an active
research towards a unified understanding of diverse topologies of networks. Such studies have
revealed that many networks with the scale-free and fractal properties exhibit the structural
multifractality, some of which are actually bifractal. Bifractality is a particular case of the
multifractal property, where only two local fractal dimensions $\dfmin$ and $\dfmax (>\dfmin$)
suffice to explain the structural inhomogeneity of a network. In this work, we investigate
analytically and numerically the multifractal property of a wide range of fractal scale-free
networks (FSFNs) including deterministic hierarchical, stochastic hierarchical, non-hierarchical,
and real-world FSFNs. Then we demonstrate how commonly FSFNs exhibit the bifractal property. The
results show that all these networks possess the bifractal nature. We conjecture from our findings
that any FSFN is bifractal. Furthermore, we find that in the thermodynamic limit the lower local
fractal dimension $\dfmin$ describes substructures around infinitely high-degree hub nodes and
finite-degree nodes at finite distances from these hub nodes, whereas $\dfmax$ characterizes local
fractality around finite-degree nodes infinitely far from the infinite-degree hub nodes. Since the
bifractal nature of FSFNs may strongly influence time-dependent phenomena on FSFNs, our results
will be useful for understanding dynamics such as information diffusion and synchronization on
FSFNs from a unified perspective.
\end{abstract}


\maketitle

\section{Introduction}
\label{sec:intro} Large-scale networks describing real-world complex systems usually have
inhomogeneous structures with large fluctuations in the degree, the number of edges incident to a
node. In fact, many complex networks exhibit the scale-free property with power-law degree
distributions \cite{Barabasi99,Albert02,Newman03}. In order to  understand properties of these
inhomogeneous networks theoretically, we need to simplify structures of networks. One such
simplification is based on an approximation that ignores degree correlations in networks. Owing to
the generating function technique, this type of approximation has successfully provided many
insights into uncorrelated complex networks \cite{Newman01}. There exist, however, networks whose
degree correlations play significant roles. For investigating such strongly correlated networks, it
is helpful to focus the fractal nature of a network. A network is fractal with the fractal
dimension $\Df$ if the minimum number of subgraphs (boxes) $N_{\text{B}}(l)$ with fixed diameter
$l$ required to cover the entire network is proportional to $l^{-\Df}$. Fractal networks are often
contrasted with small-world networks in which $N_{\text{B}}(l)$ decreases exponentially with $l$.
Since the discovery of real-world scale-free networks exhibiting the fractal nature \cite{Song05},
the fractality of complex networks has been extensively studied
\cite{Song06,Goh06,Song07,Gallos07,Kim07,Rozenfeld07,Rozenfeld10,
Watanabe15,Fujiki17a,Fujiki17b,Rosenberg20}. These studies reveal that many real networks are
fractal \cite{Kim07,Concas06,Zhang07,Kitsak07,Zhou07,Guida07}, at least, on shorter scales than the
average shortest-path distance $\langle l\rangle$ even if the networks exhibit the small-world
property on longer scales \cite{Rozenfeld10,Kawasaki10}.

If a network possesses both the scale-free and fractal properties, a single fractal dimension may
not be sufficient to fully describe the fractal nature of the network because the fractal
scale-free network (FSFN) could have a multifractal structure \cite{Furuya11}. Multifractality is a
property of inhomogeneously distributed quantities (probability measures) defined on a fractal
object \cite{Mandelbrot74}. Although the multifractal nature was initially argued for fractal
systems embedded in Euclidean space, it is possible to extend the argument to networks by replacing
Euclidean distance with shortest-path distance. In particular, if a measure is evenly assigned to
each node, but the distribution of the measure is multifractal, then the network structure itself
is considered multifractal. In recent years, numerous studies have presented various efficient
algorithms for the multifractal analysis of complex networks and examined the multifractality of
many synthetic and real-world networks by utilizing these algorithms
\cite{Rosenberg20,Wang12,Li14,Liu15,Rendon17,Rosenberg17,Pavon20,Xiao21,Ding21,Pavon22}. As an
important aspect of FSFNs, it has been shown that a specific class of FSFNs possess
\textit{bifractal} structures in which two local fractal dimensions fully characterize the fractal
nature of networks \cite{Furuya11}. More precisely, an FSFN $\G$ is always bifractal if the number
of nodes included in a super-node of the renormalized network of $\G$ is proportional to the degree
of the super-node. In fact, FSFNs formed by the $(u,v)$-flower model \cite{Rozenfeld07} and the
Song-Havlin-Makse model \cite{Song06} have been identified as bifractal networks. Since the
bifractal property of a network suggests that two qualitatively different behaviors are expected in
various local dynamics on the network \cite{Rozenfeld07,Hwang12a,Hwang12b,Hwang13}, it is
imperative to clarify to what extent FSFNs commonly exhibit bifractal structures. This, however,
remains to be revealed. Furthermore, the relationship between the two local fractal dimensions of a
bifractal network and the network structure has yet to be elucidated.

In this work, we investigate the bifractal property of broader classes of FSFNs. To this end, we
analyze four types of FSFNs, namely deterministic hierarchical FSFNs formed by the single-generator
model \cite{Yakubo22}, stochastic hierarchical FSFNs formed by the multi-generator model,
non-hierarchical FSFNs at the percolation critical points, and real-world FSFNs. The combination of
these types of networks covers a wide range of FSFNs with a variety of fractal, scale-free, and
clustering properties. Our results show that all FSFNs we examined have bifractal structures and we
conjecture that any FSFN is bifractal. Furthermore, we study how the bifractality of an FSFN
relates to its local fractal structures. In order to identify which part of the network each local
fractal dimension $\df$ describes, we compute $\df$ around each node in the FSFN by counting the
number of nodes $\tilde{\nu}_{i}(l)$ within shortest-path distance $l$ from a node $i$
\cite{Xiao21}. Our results demonstrate that there exist two distinct substructures with two local
fractal dimensions $\dfmin$ and $\dfmax$ in an infinite FSFN. The dimension $\dfmin$ describes the
fractality of a region including an infinite-degree hub node, while $\dfmax$ characterizes a region
including only finite-degree nodes.

The rest of this paper is organized as follows: Section II briefly summarizes the multifractal
property of complex networks and the possibility of structural bifractality of FSFNs found by
previous studies. Section III shows the bifractal property of various types of FSFNs including
deterministic FSFNs formed by the single-generator model, stochastic FSFNs constructed by the
multi-generator model, non-hierarchical FSFNs, and real-world FSFNs. In Sec.~IV, we clarify that
each of the two local fractal dimensions of an FSFN characterizes which substructures of the network.
We give our conclusions in Sec.~V. Hereafter, the terms \textit{distance},
\textit{diameter}, and \textit{radius} are used in the sense of shortest-path distance
(chemical distance).

\section{Multifractal property of networks}
\label{sec:2}

First we summarize the argument of the multifractal and bifractal properties of complex networks.
Consider a simple, connected network $\G$ with the set of nodes $V(\G)$ and the set of edges
$E(\G)$. Let us cover $\G$ with the minimum number of boxes (subgraphs) with a fixed diameter and
assume that a probability measure $\mu_{i}$ is defined at each node $i \in V(\G)$. The measure
$\mu_{i}$ is normalized as $\sum_{b}\sum_{i\in V(b)}\mu_{i}=1$, where $V(b)$ is the set of nodes in
a box $b$ of diameter $l$ and the first summation is taken over all the boxes. If the $q$-th moment
$Z_{q}(l)$ of the coarse-grained box measure $\mu_{b}\equiv \sum_{i\in V(b)}\mu_{i}$ satisfies
\begin{equation}
Z_{q}(l)=\sum_{b}\mu_{b}^{q} \propto l^{\tau(q)}\
\label{eq:1}
\end{equation}
and the mass exponent $\tau(q)$ is nonlinear with respect to $q$, then we say that the distribution
of $\mu_{i}$ is multifractal. Since $\tau(0)$ is identical to $-\Df$, the multifractal distribution
of $\mu_{i}$ requires the fractality of $\G$. Moreover, if the relation Eq.~(\ref{eq:1}) stands for
a constant measure $\mu_{i}=\mu_{0}$ for all $i \in V(G)$, the network structure itself is
considered multifractal. In this case, the H\"{o}lder exponent $\alpha(q)$ defined by
$\alpha(q)=d\tau(q)/dq$ is equivalent to the local fractal dimension $\df$ describing a
substructure in $\G$, because the box measure $\mu_{b}$ is proportional to $l^{\alpha(q)}$, i.e.,
$\sum_{i\in V(b)}\mu_{i}\propto l^{\alpha(q)}$, where again $l$ is the box diameter. Due to the
nonlinearity of $\tau(q)$, $\alpha(q)$ takes various values depending on $q$, so a structurally
multifractal network is, intuitively, a network in which there exist various substructures
described by different values of local fractal dimensions. Previous studies have proposed a number
of efficient algorithms for examining the multifractal nature of networks and reported the
multifractality of a wide range of synthetic and real-world fractal networks \cite{Rosenberg20,Wang12,
Li14,Liu15,Rendon17,Rosenberg17,Pavon20, Xiao21, Ding21,Pavon22}.

If the network $\G$ is not only fractal but also scale-free, $\G$ may possess a \textit{bifractal}
structure in which two local fractal dimensions suffice to characterize the fractal nature of $\G$
\cite{Furuya11}. In particular, when covering a fractal scale-free network (FSFN) $\G$ with the
minimum number of boxes, $\G$ is always bifractal if we have
\begin{equation}
\nu_{b}\propto k_{b} ,
\label{eq:2}
\end{equation}
where $\nu_{b}$ is the number of nodes in a box $b$ and $k_{b}$ is the number of neighboring
boxes of $b$ \cite{Furuya11}. The quantity $k_{b}$ is considered to be the degree of the super-node
$b$ in the renormalized network of $\G$. As shown in Appendix \ref{appendix:a}, if
Eq.~(\ref{eq:2}) holds, the mass exponent $\tau(q)$ is given by
\begin{equation}
\tau(q)=
\begin{cases}
(q-1)\Df                                                 & \text{for } q<\gamma-1 \\[4pt]
q\Df \left(\displaystyle\frac{\gamma-2}{\gamma-1}\right) & \text{for } q \ge \gamma-1
\end{cases},
\label{eq:3}
\end{equation}
where $\gamma$ is the degree exponent describing the asymptotic power-law behavior of the degree
distribution $P(k)$, such that $P(k)\propto k^{-\gamma}$ for high degree $k$. In general, the mass
exponent becomes a nonlinear function of $q$ if the measure $\mu_{i}$ varies with site $i$ and is
distributed in a multifractal manner. A typical $\tau(q)$ asymptotically approaches $\tau(q)=
\alpha_{\text{max}} q$ for $q \to -\infty$, while it approaches $\tau(q)=\alpha_{\text{min}} q$ for
$q \to \infty$. The function $\tau(q)$ describing a conventional multifractal system smoothly
(nonlinearly) connects these two asymptotic straight lines with different slopes at intermediate
$q$. Thus, the H\"{o}lder exponent takes multiple values. We emphasize that the mass exponent given
by Eq.~(\ref{eq:3}) has the same form as such a general profile of $\tau(q)$, in which the two
asymptotic regions meet each other at $q=\gamma-1$ and no intermediate region exists. This implies
that a bifractal system is a special case of a multifractal system with inhomogeneous local
fractality. From Eq.~(\ref{eq:3}), the local fractal dimension equivalent to the H\"{o}lder
exponent takes two values,
\begin{equation}
\begin{split}
\dfmax&= \Df , \\
\dfmin&= \Df \left(\displaystyle\frac{\gamma - 2}{\gamma - 1}\right) .
\end{split}
\label{eq:4}
\end{equation}
Bifractality of FSFNs has been shown \cite{Furuya11} analytically for the $(u,v)$-flower model
\cite{Rozenfeld07} and numerically for the Song-Havlin-Makse (SHM) model \cite{Song06}. The
theoretical expression of $\tau(q)$ for the $(u,v)$-flower model has often been used to benchmark
newly developed numerical algorithms for multifractal analysis
\cite{Li14,Liu15,Pavon20,Ding21,Pavon22}. However, the validity of Eqs.~(\ref{eq:3}) and
(\ref{eq:4}) has been verified only for these two kinds of FSFNs, and it remains unclear how common
the bifractal property of FSFNs is. In this work, we investigate how wide a range of FSFNs
satisfies Eq.~(\ref{eq:2}) and exhibits bifractality by examining various conceivable types of
FSFNs.

\section{Bifractality of FSFNs}
\label{sec:3}

\subsection{Deterministic hierarchical FSFNs}
\label{subsec:3-1}

\begin{figure}[tttt]
    \begin{center}
    \includegraphics[width=0.8\linewidth]{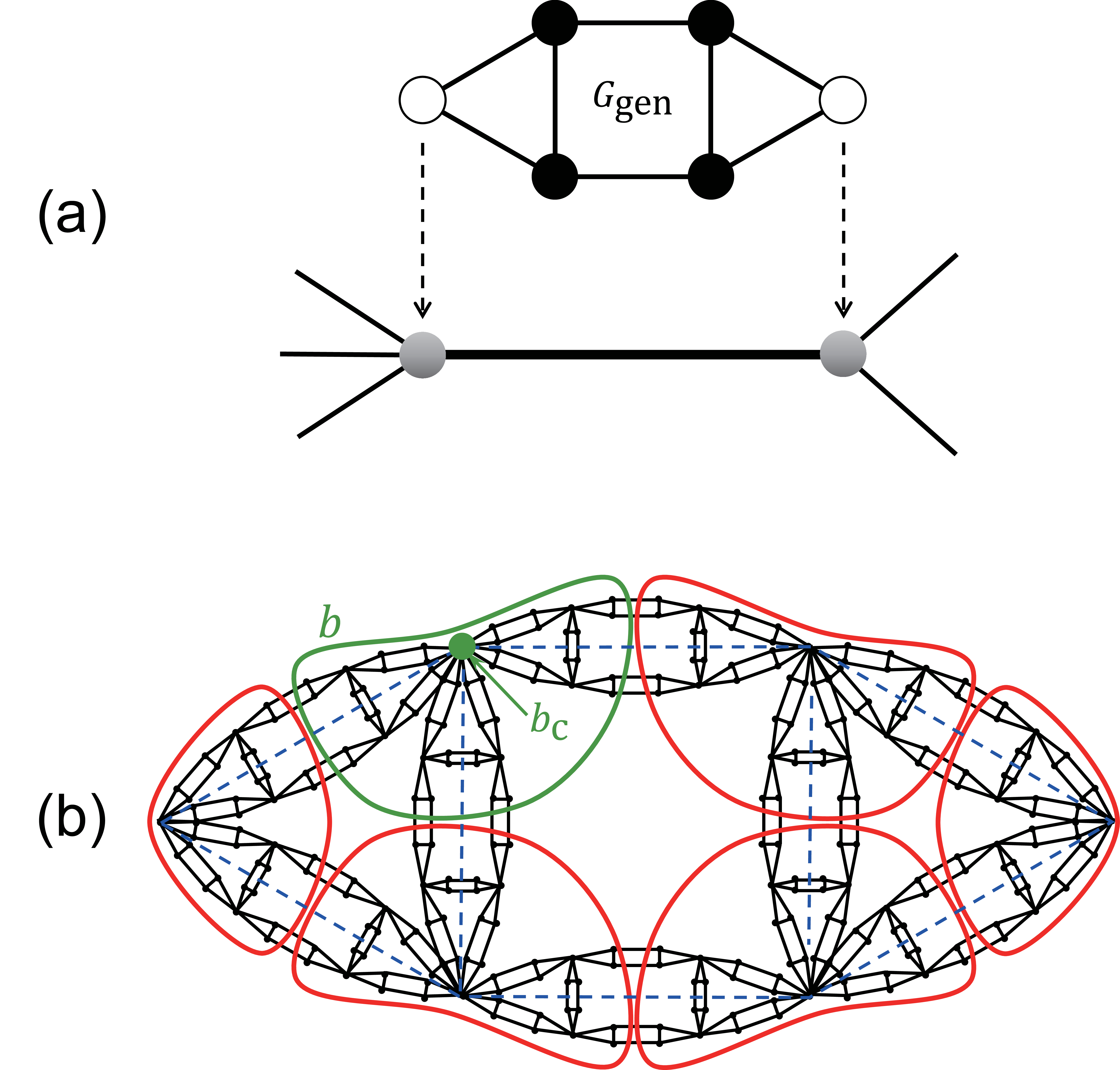}
    \caption{
(a) Example of a generator $\Ggen$ (top) and the edge replacement by $\Ggen$. Two white nodes in
$\Ggen$ represent the root nodes and four black nodes are the remaining nodes. (b) Box covering
[green (left top) and red (other) closed curves] of the third generation network $\G_{3}$ formed by
$\Ggen$ shown in (a). The six boxes constitute the network $\G_{1}$ (blue dashed lines). The green,
emphasized node is the central node of the green box $b$. The diameter ($l=9$) of a box is equal to
the diameter of $\G_{2}$.
    }
    \label{fig:1}
    \end{center}
\end{figure}
In order to examine the validity of Eq.~(\ref{eq:2}) [and thus Eqs.~(\ref{eq:3}) and (\ref{eq:4})]
for a wide range of FSFNs, we first investigate a class of deterministic hierarchical FSFNs formed
by the single-generator model \cite{Yakubo22}. In this model, the $\gen$-th generation network
$\G_{\gen}$ is recursively constructed by replacing every edge in $\G_{\gen-1}$ with a small graph
$\Ggen$ called a generator, so that two root nodes predefined in $\Ggen$ coincide with the two
terminal nodes of the replaced edge, as illustrated by Fig.~\ref{fig:1}(a). The root nodes of
$\Ggen$ must not be adjacent to each other and have degrees at least two. For simplicity, we
concentrate below on the symmetric generator in which a subgraph constructed by removing one root
node and its incident edges from $\Ggen$ has the same topology as a subgraph formed by removing
another root node and its incident edges from $\Ggen$. It is easy to extend our argument below to
the case of an asymmetric generator. We initialize the network $\G_{0}$ with a single edge with two
terminal nodes, so the first generation network $\G_{1}$ is $\Ggen$ itself.

Let us consider a generator $\Ggen$ with $m_{\text{gen}}$ edges, in which the two root nodes have
degree $\kappa$ and they are separated by distance $\lambda$. The number of nodes in the $\gen$-th
generation network $\G_{\gen}$ by the single-generator model is \cite{Yakubo22}
\begin{equation}
N_{\gen}=2+\frac{n_{\text{rem}}(m_{\text{gen}}^{\gen}-1)}{m_{\text{gen}}-1} ,
\label{eq:5}
\end{equation}
where $n_{\text{rem}}$ is the number of ``remaining nodes" defined as nodes other than the root
nodes in the generator $\Ggen$. For $\gen\gg 1$, $\G_{\gen}$ becomes fractal and scale-free. The
degree exponent $\gamma$ and fractal dimension $\Df$ are given by
\begin{gather}
\gamma=1+\frac{\log m_{\text{gen}}}{\log \kappa} \label{eq:6} \\
\Df=\frac{\log m_{\text{gen}}}{\log \lambda} \label{eq:7} .
\end{gather}
In addition to $\gamma$ and $\Df$, various properties of $\G_{\gen}$, such as the clustering
coefficient, degree correlation, and percolation properties, can be analytically obtained by using
the quantities determined by the structure of the generator \cite{Yakubo22}. We can obtain a wide
range of deterministic hierarchical FSFNs, including the $(u,v)$-flower and those by the SHM model,
by choosing from a variety of generators.

Here, we show that Eq.~(\ref{eq:2}) holds for FSFNs formed by this single-generator model. To this
end, let us cover the $\gen$-th generation network $\G_{\gen}$ with boxes of fixed diameter $l$
so that the boxes constitute the network $\G_{\gen'}$ as shown by Fig.~\ref{fig:1}(b), where
$0 \le \gen' \le \gen$. Since the boxes in this covering scheme correspond to the \textit{super-nodes}
of the renormalized network $\G_{\gen'}$, the number of boxes $N_{\text{B}}(l)$ coincides with the
number of nodes in $\G_{\gen'}$, namely $N_{\gen'}$ given by Eq.~(\ref{eq:5}). While it is not
obvious whether $N_{\text{B}}(l)(=N_{\gen'})$ provides the minimum number of covering boxes for
$\G_{\gen}$, $N_{\text{B}}(l)$ would be at least proportional to the minimum number, because we can
easily show the relation $N_{\text{B}}(l)\propto l^{-\Df}$ with the fractal dimension $\Df$ given
by Eq.~(\ref{eq:7}).

In order to count the number of nodes $\nu_{b}$ in a box $b$ under this box-covering scheme, we
define the central node $b_{\text{c}}$ of the box $b$ as the node in $\G_{\gen}$ corresponding to
the super-node $b$ in $\G_{\gen'}$. In Fig.~\ref{fig:1}(b), the green node is the central node of
the green box $b$. The quantity $\nu_{b}$ is the number of nodes whose nearest central node is
$b_{\text{c}}$. We remark that a super-edge [a blue dashed line in Fig.~\ref{fig:1}(b)] of the
renormalized network $\G_{\gen'}$ represents $\G_{\gen-\gen'}$ having $N_{\gen-\gen'}$ nodes and
the half of these nodes are contained in the box $b$. More precisely, the half of the nodes in
$\G_{\gen-\gen'}$ other than the two central nodes contribute to $\nu_{b}$ from a single
super-edge. Note that for any node that is equidistant from two central nodes, we consider that a
half of the node belongs to one box while the other half belongs to another. Such a treatment is
justified in the mean-field sense. Denoting the degree of the super-node $b$ as $k_{b}$, $\nu_{b}$
is given by $k_{b}(N_{\gen-\gen'}-2)/2+1$, where ``$+1$" is the contribution from the central node
itself. Therefore, using Eq.~(\ref{eq:5}), we obtain,
\begin{equation}
\nu_{b}=1+k_{b}\frac{n_{\text{rem}}(m_{\text{gen}}^{\gen-\gen'}-1)}{2(m_{\text{gen}}-1)} .
\label{eq:8}
\end{equation}
This relation can be also explained by counting the number of nodes, that appear in the growth
process from $\G_{\gen'}$ to $\G_{\gen}$, whose nearest node belonging to $\G_{\gen'}$ is the
super-node $b$. The second term of Eq.~(\ref{eq:8}) dominates the first term for $\gen\gg \gen'$,
thus we have Eq.~(\ref{eq:2}) for the network $\G_{\gen}$ if the box size is much smaller than the
size of $\G_{\gen}$.

\begin{figure}[tttt]
    \begin{center}
    \includegraphics[width=0.9\linewidth]{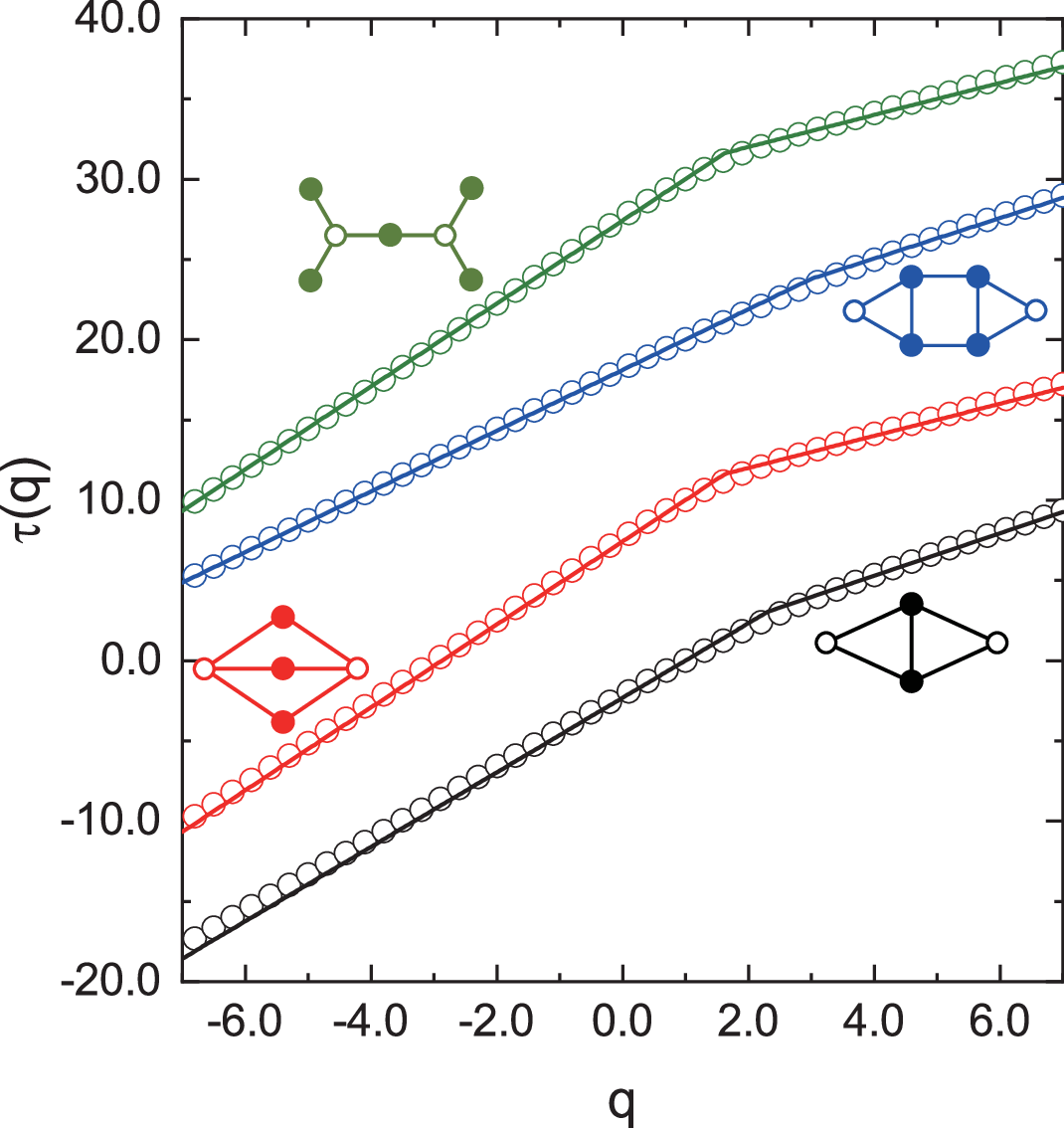}
    \caption{
Mass exponents $\tau(q)$ for deterministic hierarchical FSFNs generated by the single-generator
model. Solid lines represent $\tau(q)$'s calculated theoretically by Eq.~(\ref{eq:9}) and symbols
indicate those computed numerically by using the sandbox algorithm \cite{Liu15}. Different colors
stand for results for different FSFNs. For graphical reasons, the red (second from the bottom),
blue (second from the top), and green (top) results are shifted upward by $10$, $20$, and $30$,
respectively. The generators employed for our calculations are indicated by small networks in the
same colors. Open nodes represent the root nodes. For numerical calculations, the 7th, 6th, 5th,
and 6th generation networks formed by the black, red, blue, and green generators are used, which
contain $39064$ ($78125$), $27995$ ($46656$), $18726$ ($32768$), and $46657$ ($46656$) nodes
(edges), respectively.
    }
    \label{fig:2}
    \end{center}
\end{figure}

Since Eq.~(\ref{eq:2}) holds for any $\G_{\gen}$ with $\gen\gg\gen'\gg 1$, deterministic
hierarchical FSFNs formed by the single-generator model exhibit bifractal structures. Using
Eqs.~(\ref{eq:3}), (\ref{eq:6}), and (\ref{eq:7}), the mass exponent $\tau(q)$ is given by
\begin{equation}
\tau(q)=
\begin{cases}
    q \displaystyle\frac{\log (m_{\text{gen}}/\kappa)}{\log \lambda}  & \text{for } q \ge \displaystyle\frac{\log m_{\text{gen}}}{\log \kappa} \\[10pt]
    (q-1) \displaystyle\frac{\log m_{\text{gen}}}{\log \lambda}       & \text{for } q < \displaystyle\frac{\log m_{\text{gen}}}{\log \kappa}
\end{cases},
\label{eq:9}
\end{equation}
and it follows from Eq.~(\ref{eq:4}) that the local fractal dimensions are
$\dfmax=\log m_{\text{gen}}/\log \lambda$ and $\dfmin=\log(m_{\text{gen}}/\kappa)/\log \lambda$.
To confirm the validity of our theoretical prediction, we have calculated numerically $\tau(q)$
for various FSFNs formed by different generators by applying the sandbox algorithm\cite{Liu15}.
As shown in Fig.~\ref{fig:2}, numerical results agree well with the mass exponent presented by
Eq.~(\ref{eq:9}).

\subsection{Stochastic hierarchical FSFNs}
\label{subsec:3-2}

While it has been established that numerous hierarchical FSFNs generated by deterministic
algorithms display bifractal structures, FSFNs do not necessarily have deterministic structures. In
order to examine the bifractality of stochastic hierarchical FSFNs, we extend the single-generator
model to a model with stochastic edge replacements by multiple generators. In this multi-generator
model, we prepare a set of $s$ generators $\{\Ggen^{(1)}, \Ggen^{(2)}, \dots, \Ggen^{(s)}\}$.
As in the case of the single-generator model, two non-adjacent root nodes are specified in advance in
each generator, where the degrees of the root nodes are $2$ or higher. For simplicity, these
generators are assumed to be symmetric. To construct the $\gen$-th generation network $\G_{\gen}$,
we replace every edge in the previous generation network $\G_{\gen-1}$ with one of the multiple
generators $\Ggen^{(i)}$ with the predefined probability $p^{(i)}$, where
$\sum_{i=1}^{s}p^{(i)}=1$. The edge replacement method is the same as that in the single-generator
model.

As shown in Appendix \ref{appendix:b}, the number of nodes $N_{\gen}$ in the $\gen$-th generation
network $\G_{\gen}$ is
\begin{equation}
N_{\gen}=2+\frac{\nremav\left(\mgenav^{\gen}-1\right)}{\mgenav-1} ,
\label{eq:10}
\end{equation}
where $\mgenav=\sum_{i=1}^{s}m_{\text{gen}}^{(i)}p^{(i)}$ and
$\nremav=\sum_{i=1}^{s}n_{\text{rem}}^{(i)}p^{(i)}$ are the mean number of edges and the mean
number of remaining nodes in a generator averaged over the multiple generators, respectively. Here,
$m_{\text{gen}}^{(i)}$ and $n_{\text{rem}}^{(i)}$ denote respectively the number of edges and the
number of remaining nodes in $\Ggen^{(i)}$. Appendix \ref{appendix:b} also shows that networks
constructed by the multi-generator model exhibit the scale-free and fractal properties as in the case
of the single-generator model. The degree exponent $\gamma$ and fractal dimension $\Df$ are given
by
\begin{gather}
\gamma=1+\frac{\log \mgenav}{\log \kappaav} ,      \label{eq:11} \\
\Df=\frac{\log \mgenav}{\log \lambdaav} , \label{eq:12}
\end{gather}
where $\kappaav=\sum_{i=1}^{s}\kappa^{(i)}p^{(i)}$ is the mean degree of a root node averaged over
the multiple generators, and $\kappa^{(i)}$ is the degree of the root node in generator
$\Ggen^{(i)}$. In addition, $\lambdaav$ represents the mean inter-root-node distance. This quantity
is determined by the set of distances $\{\lambda^{(i)}:1\le i \le s\}$ between the root nodes in
the multiple generators, but not the simple average of $\lambda^{(i)}$. Appendix \ref{appendix:b}
shows how $\lambdaav$ is calculated from $\{\lambda^{(i)}\}$. It should be emphasized that the
multi-generator model encompasses a much broader class of FSFNs than that of the single-generator
model because $\mgenav$, $\kappaav$, and $\lambdaav$ can take non-integer values.

We cover the network $\G_{\gen}$ with boxes of a fixed diameter as we did for the single-generator
model. Namely, $\G_{\gen}$ is covered so that the covering boxes constitute the network
$\G_{\gen'}$ with $\gen'\ll \gen$, where $\G_{\gen'}$ is a network appearing in the stochastic
construction process from $\G_{0}$ to $\G_{\gen}$. A box $b$ contains half of the nodes within the
subgraphs constituting $k_{b}$ super-edges from the super-node $b$ in the renormalized network
$\G_{\gen'}$, as in the single-generator model, but these super-edges stand for different
realizations of $\G_{\gen-\gen'}$ in this stochastic model. However, if $\gen\gg \gen'\gg 1$, i.e.,
the super-degree $k_{b}$ and sizes of networks $\G_{\gen-\gen'}$ are large enough, then we can
approximate the number of nodes $\nu_{b}$ in the box $b$ as $\nu_{b}=k_{b}(N_{\gen-\gen'}-2)/2+1$,
where $N_{\gen-\gen'}$ is given by Eq.~(\ref{eq:10}). This leads to Eq.~(\ref{eq:8}) for $\nu_{b}$
with $n_{\text{rem}}$ and $m_{\text{gen}}$ replaced by $\nremav$ and $\mgenav$, respectively. Since
Eq.~(\ref{eq:2}) holds when $N_{\gen-\gen'}\gg 1$ for $\gen\gg \gen'$, we can conclude that
stochastic hierarchical FSFNs formed by the multi-generator model possess bifractal structures.
Using Eqs.~(\ref{eq:3}), (\ref{eq:11}), and (\ref{eq:12}), the mass exponent $\tau(q)$ is given by
\begin{equation}
\tau(q)=
\begin{cases}
    q \displaystyle\frac{\log (\mgenav/\kappaav)}{\log \lambdaav}  & \text{for } q \ge \displaystyle\frac{\log \mgenav}{\log \kappaav} \\[10pt]
    (q-1) \displaystyle\frac{\log \mgenav}{\log \lambdaav}         & \text{for } q < \displaystyle\frac{\log \mgenav}{\log \kappaav}
\end{cases},
\label{eq:13}
\end{equation}
and it follows from Eq.~(\ref{eq:4}) that the two local fractal dimensions are
$\dfmax=\log\mgenav/\log\lambdaav$ and $\dfmin=\log(\mgenav/\kappaav)/\log\lambdaav$.

\begin{figure}[tttt]
    \begin{center}
    \includegraphics[width=0.9\linewidth]{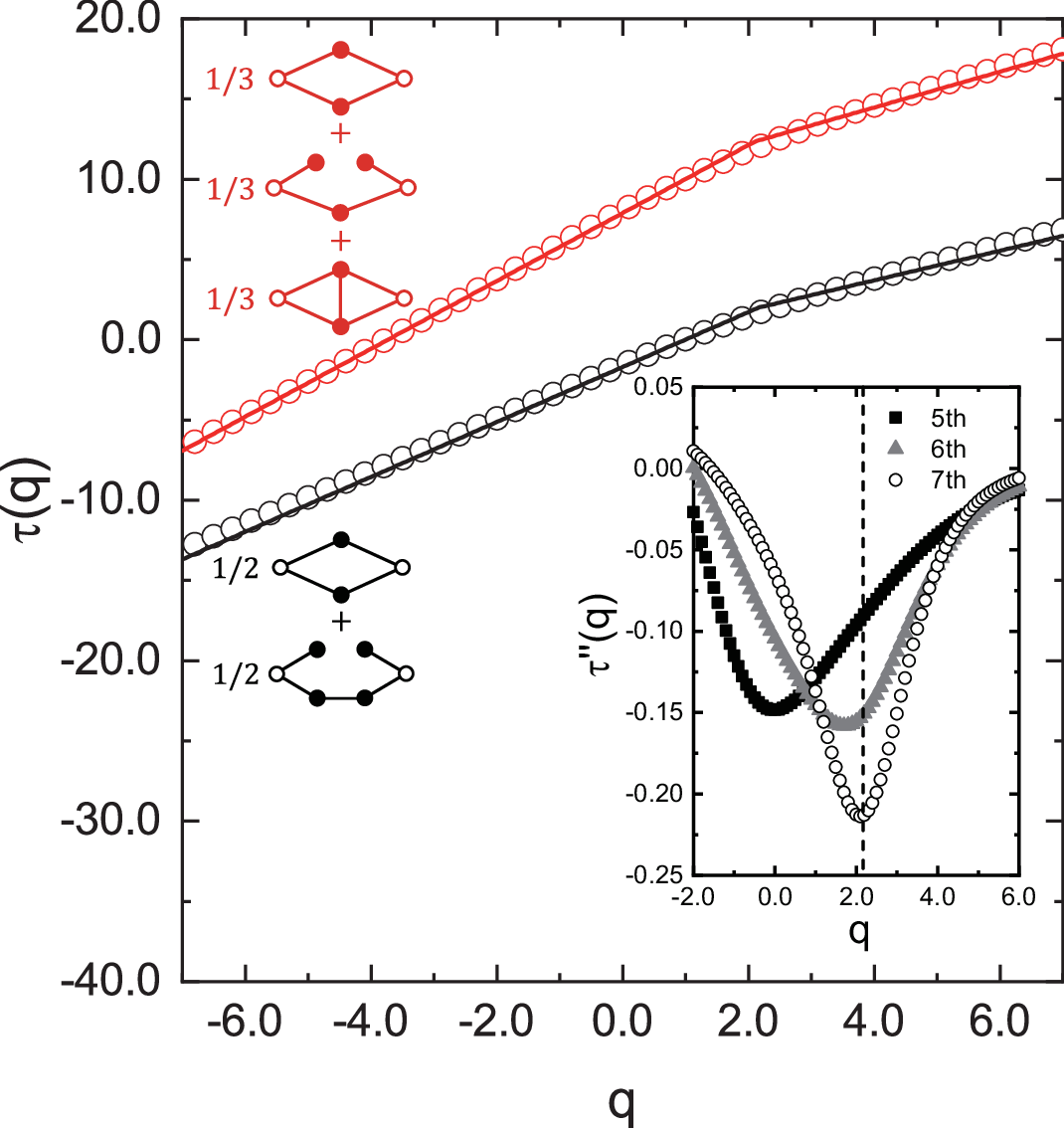}
    \caption{
Mass exponents $\tau(q)$ for stochastic hierarchical FSFNs generated by the multi-generator model.
Solid lines represent $\tau(q)$'s calculated theoretically by Eq.~(\ref{eq:13}) and symbols
indicate those computed numerically for 7th generation networks averaged over $100$ [black (bottom)
results] or $50$ [red (top) results] realizations by using the sandbox algorithm \cite{Liu15}.
Different colors stand for results for different combinations of generators indicated by small
networks in the same colors. Fractions beside the generators represent the edge replacement
probabilities. For graphical reasons, the red results are shifted upward by $10$. The inset shows
the numerically calculated second derivatives of $\tau(q)$ for 5th, 6th, and 7th generation FSFNs
formed by the combination of black generators. The vertical dashed line indicates the folding point
of $\tau(q)$ that is theoretically predicted.
    }
    \label{fig:3}
    \end{center}
\end{figure}
The above theoretical prediction has been confirmed numerically by computing $\tau(q)$ for FSFNs
formed by the multi-generator model with two and three generators. Our results are shown in
Fig.~\ref{fig:3}.  Although the numerical results for 7th generation networks deviate slightly from
$\tau(q)$ presented by Eq.~(\ref{eq:13}), they are basically in agreement with the theoretical
results. Such deviations are caused by finite-size effects. As expected from our analytical
argument, Eq.~(\ref{eq:2}) holds in the multi-generator model when $\gen\gg\gen'\gg1$ and the
bifractal behavior of $\tau(q)$ is realized in large FSFNs. The sizes of FSFNs employed for
numerical calculations (on average, $N_{7}=32030$ and $20085$ for black and red results in
Fig.~\ref{fig:3}, respectively) may not be large enough. To investigate the finite-size effect, we
calculate the $q$-dependence of the second derivative $\tau^{\prime\prime}(q)$, namely the
curvature of $\tau(q)$. The results shown in the inset of Fig.~\ref{fig:3} demonstrate that as the
generation of FSFNs increases, $\tau(q)$ begins to fold abruptly and the folding point approaches
the theoretical value $q=\log\mgenav/\log\kappaav$. While Fig.~\ref{fig:3} only shows
$\tau^{\prime\prime}(q)$ for the combination of black generators, we obtained similar results also
for the red generator combination. Furthermore, Eq.~(\ref{eq:13}) has been confirmed for various
combinations of generators other than those shown in Fig.~\ref{fig:3}.

\subsection{Non-hierarchical FSFNs}
\label{subsec:3-3}

Structures of networks treated so far are, by construction, hierarchical. Do FSFNs with less clear
hierarchical structures also exhibit the bifractal property? To answer this question, we
investigate giant connected components of uncorrelated scale-free networks at their percolation
critical points. If the degree distribution of a scale-free random network (SFRN) $\G$ at
criticality is given by $P(k)\propto k^{-\gamma}$ for high degree $k$, then the degree distribution
$P_{\text{GC}}(k)$ of the giant component $\G_{\text{GC}}$ of $\G$ follows
$P_{\text{GC}}(k)\propto k^{-\gamma'}$ with $\gamma'=\gamma-1$ \cite{Bialas08}. In addition,
$\G_{\text{GC}}$ takes a fractal structure with the fractal dimension $\Df=2$ for $\gamma\ge 4$ or
$\Df=(\gamma-2)/(\gamma-3)$ for $3<\gamma<4$ \cite{Cohen03}. These facts imply that
$\G_{\text{GC}}$ is an FSFN.

Let us cover the giant component $\G_{\text{GC}}$ with boxes, each of which consists of nodes
within distance $l$ from a central node of degree $k$, and count the number of nodes $\nu_{b}$ in
a box $b$. We should note here that $\G_{\text{GC}}$ exhibits a long-range degree correlation
though $\G$ is uncorrelated. In general, long-range degree correlations of networks are described
by the conditional probability $P(k,k'|l)$ of randomly chosen two nodes separated by $l$ from each
other having degrees $k$ and $k'$ or the probability $P(k'|k,l)$ that a node separated by $l$ from
a randomly chosen node of degree $k$ has degree $k'$ \cite{Fujiki18}. These probabilities for the
giant component of a random network, $P_{\text{GC}}(k,k'|l)$ and $P_{\text{GC}}(k'|k,l)$, have been
studied by Mizutaka and Hasegawa \cite{Mizutaka20} for the case that the third moment
$\sum_{k}k^{3}P(k)$ is finite. In particular, the probability $P_{\text{GC}}(k'|k,l)$ is given by
\begin{equation}
P_{\text{GC}}(k'|k,l)=\frac{1-v^{l-1}u^{k+k'-2}}{1-v^{l-1}u^{k}}\frac{k'P(k')}{z_{1}} ,
\label{eq:14}
\end{equation}
where $u$ is the solution of $u=G_{1}(u)$ with the generating function $G_{1}(u)$ defined by
$G_{1}(u)=\sum_{k}kP(k)u^{k-1}/z_{1}$, $v=G'_{1}(u)/G'_{1}(1)$, and $z_{1}=\sum_{k}kP(k)$. Since
$u$ is the probability that an edge does not lead to the giant component, we have $u<1$ in the
percolating phase and $u=1$ at the critical point. Assuming that $u=1-\varepsilon$ with a small
positive quantity $\varepsilon$ near the critical point, we have $v=1-(z_{3}/z_{2})\varepsilon$,
where $z_{2}=\sum_{k}k(k-1)P(k)$ and $z_{3}=\sum_{k}k(k-1)(k-2)P(k)$. Substituting these
expressions of $u$ and $v$ into Eq.~(\ref{eq:14}) and taking the limit of $\varepsilon\to 0$,
$P_{\text{GC}}(k'|k,l)$ at criticality is given by
\begin{equation}
P_{\text{GC}}(k'|k,l)=\frac{z_{3}(l-1)+z_{2}(k+k'-2)}{z_{3}(l-1)+z_{2}k}\frac{k'P(k')}{z_{1}} ,
\label{eq:15}
\end{equation}
if $z_{3}$ is finite. We utilize this probability to examine analytically the validity of
Eq.~(\ref{eq:2}) for $\G_{\text{GC}}$ at criticality under the condition of finite $z_{3}$ (i.e.,
$\gamma>4$).

In order to calculate the number of nodes $\nu_{b}$ in the box $b$, we first consider the number of
nodes of degree $k'$ at distance $l$ from a node $i_{k}$ of degree $k$. Denote this quantity by
$n(k'|k,l)$. Assuming that $\G_{\text{GC}}$ at the critical point has a tree structure, $n(k'|k,l)$
satisfies the following relation:
\begin{equation}
n(k'|k,l)=P_{\text{GC}}(k'|k,l)\sum_{k''}(k''-1)n(k''|k,l-1) .
\label{eq:16}
\end{equation}
In this recurrence relation, the sum in the right-hand side represents the number of nodes at
distance $l$ from the node $i_{k}$. The quantity $n(k'|k,l)$ at $l=1$ is obviously given by
$n(k'|k,l=1)=kP_{\text{GC}}(k'|k,l=1)$. Thus, using Eq.~(\ref{eq:15}), we have
\begin{equation}
n(k'|k,l=1)=\frac{1}{z_{1}}(k+k'-2)k'P(k') .
\label{eq:17}
\end{equation}
Under this initial condition, the solution of the recurrence equation Eq.~(\ref{eq:16}) reads
\begin{equation}
n(k'|k,l)=\frac{1}{z_{1}^{2}}\left[ (l-1)z_{3}+(k+k'-2)z_{1} \right]k'P(k') .
\label{eq:18}
\end{equation}
Here we used the equality $z_{1}=z_{2}$ at the critical point \cite{{Molloy95}}. Since the number
of nodes $n_{l}(k)$ at distance $l$ from the node $i_{k}$ of degree $k$ is given by
$n_{l}(k)=\sum_{k'}n(k'|k,l)$, it follows from Eq.~(\ref{eq:18}) that
\begin{eqnarray}
n_{l}(k)=\frac{1}{z_{1}}\left[(l-1)z_{3}+kz_{1}\right] .
\label{eq:19}
\end{eqnarray}
Therefore, the number of nodes $\nu_{b(l)}$ in the box $b(l)$ consisting of nodes within distance
$l$ from the central node $i_{k}$ of degree $k$, given by $\nu_{b(l)}=1+\sum_{l'=1}^{l}n_{l'}(k)$,
is expressed as
\begin{equation}
\nu_{b(l)}=1+\frac{z_{3}}{2z_{1}}l(l-1)+kl .
\label{eq:20}
\end{equation}
On the other hand, the number of neighboring boxes $k_{b(l)}$ of $b(l)$ is equivalent to
$n_{l+1}(k)$ under the tree approximation for $\G_{\text{GC}}$, because each of the nodes at
distance $l+1$ from node $i_{k}$ belongs to a separate neighboring box. Hence, Eq.~(\ref{eq:19})
gives
\begin{eqnarray}
k_{b(l)}=\frac{z_{3}}{z_{1}}l+k .
\label{eq:21}
\end{eqnarray}
When we cover a huge scale-free $\G_{\text{GC}}$ with the minimum number of boxes, the degree $k$
of the central node of a box becomes quite large, and we can approximate Eqs.~(\ref{eq:20}) and
 (\ref{eq:21}) as $\nu_{b(l)}\simeq kl$ and $k_{b(l)}\simeq k$. Therefore, Eq.~(\ref{eq:2})
holds also for the giant component $\G_{\text{GC}}$ of an SFRN at the percolation critical point,
and we can conclude that $\G_{\text{GC}}$, an example of non-hierarchical FSFNs, shows the
bifractal property.

The above analytical argument is valid only for $\gamma>4$, but there is no obvious reason that
$\G_{\text{GC}}$ is not bifractal for $\gamma\le 4$. Thus, it is plausible that the giant component
in a critical SFRN is bifractal independently of $\gamma$. If this is the case, considering that
$P_{\text{GC}}(k)\propto k^{-\gamma'}$ with $\gamma'=\gamma-1$ and $\Df=2$ for $\gamma\ge 4$ or
$\Df=(\gamma-2)/(\gamma-3)$ for $3<\gamma<4$, the mass exponent $\tau(q)$ for $\G_{\text{GC}}$ is
given by
\begin{equation}
\tau(q)=
\begin{cases}
  q                                           & \text{for } q\ge \gamma-2\\[8pt]
  \displaystyle(q-1)\frac{\gamma-2}{\gamma-3} & \text{for } q<\gamma-2
\end{cases},
\label{eq:22}
\end{equation}
for $3<\gamma<4$, or
\begin{equation}
\tau(q)=
\begin{cases}
  2q \left(\displaystyle\frac{\gamma-3}{\gamma-2}\right) & \text{for } q\ge \gamma-2 \\[8pt]
  2(q-1)                                                 & \text{for } q<\gamma-2
\end{cases},
\label{eq:23}
\end{equation}
for $\gamma\ge 4$. The local fractal dimensions are then $\dfmax=(\gamma-2)/(\gamma-3)$ and
$\dfmin=1$ for $3<\gamma<4$ or $\dfmax=2$ and $\dfmin=2(\gamma-3)/(\gamma-2)$ for $\gamma\ge 4$.

\begin{figure}[tttt]
    \begin{center}
    \includegraphics[width=0.9\linewidth]{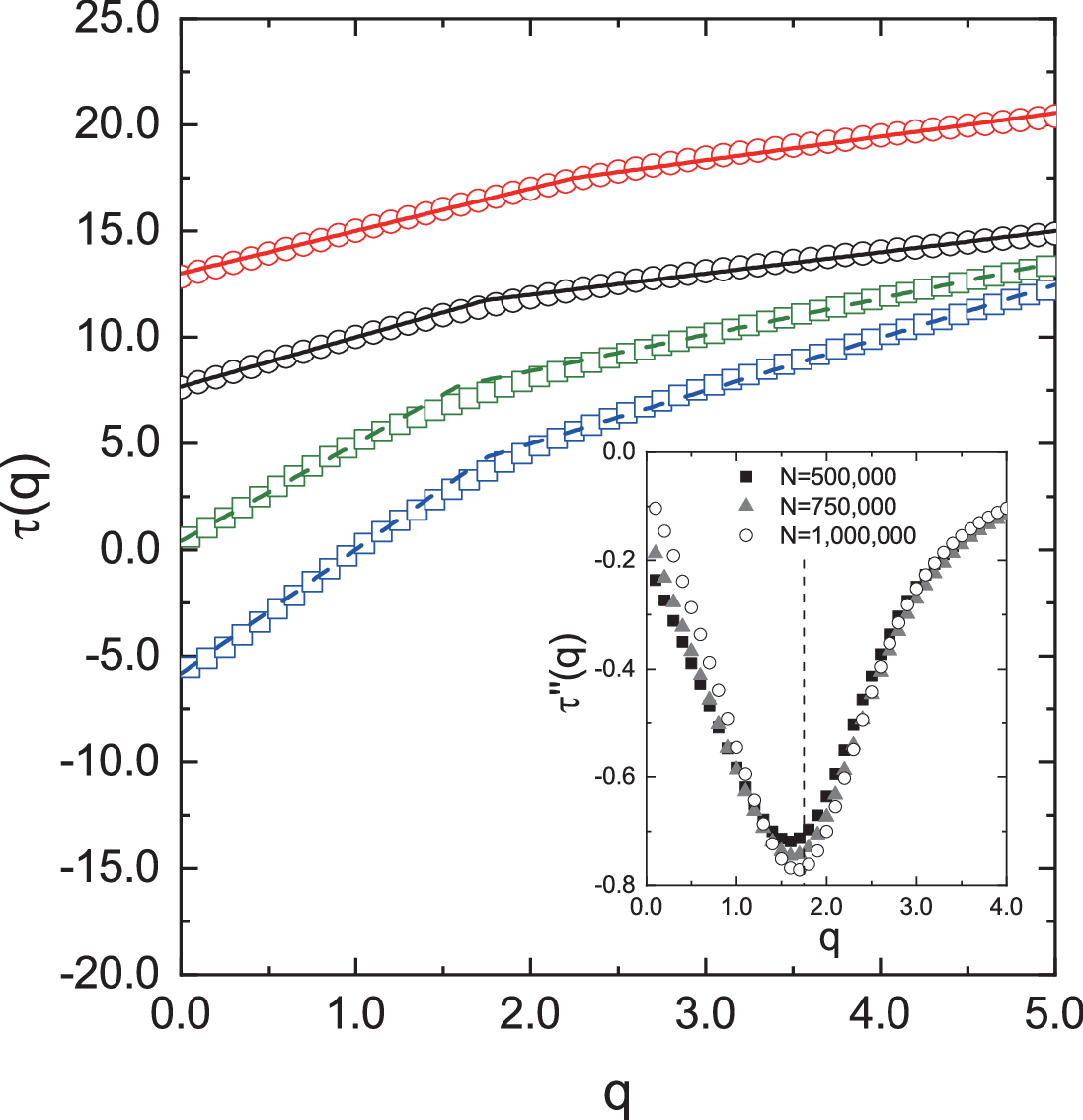}
    \caption{
Mass exponents $\tau(q)$ for non-hierarchical FSFNs (open circles and solid lines) and for
real-world FSFNs (open squares and dashed lines). Solid lines represent $\tau(q)$'s calculated
theoretically for the giant components in SFRNs at the percolation critical point, which are given
by Eq.~(\ref{eq:22}) or (\ref{eq:23}). Open circles indicate those computed numerically by using
the sandbox algorithm \cite{Liu15} for the giant components in critical SFRNs ($N=10^{6}$) formed
by the configuration model. Black (second from the top) and red (top) data show results for the
degree exponents $\gamma=3.75$ and $4.25$, respectively. Numerical results are averaged over $100$
realizations and the average giant component sizes are $12829.8$ for $\gamma=3.75$ and $14631.4$
for $\gamma=4.25$. The inset shows the numerically calculated second derivatives of $\tau(q)$ for
the giant components in critical SFRNs with $5.0\times 10^{5}$, $7.5\times 10^{5}$, and $10^{6}$
nodes and for $\gamma=3.75$. The vertical dashed line indicates the folding point of $\tau(q)$ that
is theoretically predicted. Blue (bottom) and green (second from the bottom) squares in the main
panel indicate the mass exponents $\tau(q)$ for the World-Wide Web and protein interaction network
calculated by the sandbox algorithm. Colored dashed lines represents $\tau(q)$ given by
Eq.~(\ref{eq:3}) with $\gamma$ and $\Df$ measured for these real-world networks. For graphical
reasons, the red, blue, and green results are shifted upward by $10$, $20$, and $30$, respectively.
    }
    \label{fig:4}
    \end{center}
\end{figure}

To confirm the above arguments numerically, we prepare SFRNs formed by the configuration model
\cite{Newman01,Bender78}. The degree distribution is chosen as
\begin{equation}
P(k)=\frac{c}{k^{\gamma}+d^{\gamma}} ,
\label{eq:24}
\end{equation}
for $1\le k\le 1000$ and $P(0)=P(k>1000)=0$, which is proportional to $k^{-\gamma}$ for $k\gg d$.
The parameter $d$ can control the moments of $P(k)$ and $c$ is the normalization constant. The
value of $d$ for a given $\gamma$ is chosen so that the critical condition $\langle
k^{2}\rangle/\langle k\rangle=2$ \cite{{Molloy95}} is satisfied. We employed two values of
$\gamma$, i.e., $\gamma=4.25$ and $3.75$. The parameter $d$ for these values of $\gamma$ takes
$d=1.534$ (for $\gamma=4.25$) and $1.075$ (for $\gamma=3.75$). The giant components in these
networks are fractal with the fractal dimensions $\Df=2$ (for $\gamma=4.25$) and $2.33$ (for
$\gamma=3.75$). As shown in Fig.~\ref{fig:4}, the mass exponents $\tau(q)$ for the giant components
in these SFRNs support the theoretical prediction for both $\gamma>4$ and $\gamma\le4$. As in the
case of stochastic hierarchical FSFNs, the inset in Fig.~\ref{fig:4} shows that the deviation from
the theoretical prediction due to the finite-size effect decreases as the system size increases.

\subsection{Real-world FSFNs}
\label{subsec:3-4}

Our arguments so far demonstrate that many types of FSFNs possess bifractal structures, which leads
us to suspect that any FSFN is bifractal. This is plausible also from the following considerations.
Assume that $\G$ is an FSFN. When $\G$ is covered by boxes of diameter $l$, these boxes and
connections between them construct a renormalized network $\G'$. As in the case of deterministic
hierarchical FSFNs, we can define the central node $b_{\text{c}}$ of a box $b$ as the node in $\G$
corresponding to the super-node $b$ in $\G'$. Due to the fractal nature of $\G$, the renormalized
network $\G'$ can be regarded as a network with these central nodes and equivalent (in a
statistical sense) super-edges connecting central nodes [see Fig.~\ref{fig:1}(b)]. Thus, the
original subgraphs in $\G$ corresponding to these super-edges have almost the same number of nodes
$n$. If the degree of the super-node $b$ is $k_{b}$ in $\G'$, $k_{b}$ super-edges are connected to
the central node $b_{\text{c}}$. The number of nodes $\nu_{b}$ in the box $b$ is then given by
$\nu_{b}\simeq nk_{b}/2$, which realizes Eq.~(\ref{eq:2}). Therefore, we can expect that any kind
of FSFN exhibits a bifractal structure.

To check the above conjecture, two kinds of real-world FSFNs have been numerically analyzed. One is
the World-Wide Web (WWW) of University of Notre Dame\cite{Albert99} with $325729$ nodes and the
other is the largest connected component of the human protein interaction network
(PIN)\cite{Ewing07} with $2217$ nodes. Data for both networks are freely available at
``Netzschleuder network catalogue and repository"\cite{Netzschleuder}. Treating the WWW as
undirected, we computed the degree distributions $P(k)$ for these networks and observed the
power-law behaviors of $P(k)$ with the exponents $\gamma=2.75\pm 0.08$ and $2.59\pm 0.22$ for the
WWW and PIN, respectively. To ensure consistency with the multifractal analysis, we measured the
fractal dimensions of these networks using the sandbox method, and found $\Df=5.82\pm0.10$ for the
WWW and $4.60\pm0.60$ for the PIN. As shown in Fig.~\ref{fig:4}, the mass exponents predicted by
Eq.~(\ref{eq:3}) with these observed values of $\gamma$ and $\Df$ well reproduce $\tau(q)$
calculated numerically for these real networks. These results support our conjecture that any FSFN
including real-world networks is bifractal.

\section{Local fractal dimension}
\label{sec:chap4}

The bifractal property of an FSFN implies that there exist two local fractal dimensions in the
network. It is, however, unclear which substructures of the network are characterized by $\dfmin$
and $\dfmax$. To clarify the correspondence between the substructures and the two local fractal
dimensions, we examine the local structure centered around each node $i$ in a $\gen$-th generation
network $\G_{\gen}$ formed by the single-generator model argued in Sec.~\ref{subsec:3-1}. Let node
$i$ be one of the nodes that first appears in the $\gen_{0}$-th generation ($1\le \gen_{0}\ll\gen$),
and $k$ be the initial degree of this node when it first appears. Then, the degree $k_{\gen}$ of
the node $i$ in $\G_{\gen}$ is given by
\begin{equation}
k_{\gen}=k\kappa^{\gen-\gen_{0}} ,
\label{eq:25}
\end{equation}
where $\kappa$ is the degree of the root node in the generator $\Ggen$. Since we assume $\gen\gg\gen_{0}$,
the node $i$ is a hub node with high degree $k_{\gen}$. Now, we count the number of nodes
$\tilde{\nu}_{\gen'}$ in $\G_{\gen}$ within distance $L_{\gen'}$ from the node $i$, where
$L_{\gen'}$ is the diameter of the $\gen'$-th generation network $\G_{\gen'}$ with $1\ll \gen'\le
\gen-\gen_{0}$. Since there exist $k\kappa^{\gen-\gen_{0}-\gen'}$ subgraphs $\G_{\gen'}$ within
distance $L_{\gen'}$ from the node $i$, we have
\begin{eqnarray}
\tilde{\nu}_{\gen'}&=&(N_{\gen'}-1)k\kappa^{\gen-\gen_{0}-\gen'}+1 \nonumber \\
&\simeq& N_{\gen'}k\kappa^{\gen-\gen_{0}-\gen'}
\label{eq:26}
\end{eqnarray}
where $N_{\gen'}$ is the number of nodes in $\G_{\gen'}$. Equation (\ref{eq:5}) for $N_{\gen'}$
with $\gen'\gg 1$ gives
\begin{equation}
N_{\gen'}\simeq m_{\text{gen}}N_{\gen'-1} .
\label{eq:27}
\end{equation}
Thus, it follows from Eqs.~(\ref{eq:26}) and (\ref{eq:27}) that
\begin{equation}
\tilde{\nu}_{\gen'}=\left(\frac{m_{\text{gen}}}{\kappa}\right)\tilde{\nu}_{\gen'-1} .
\label{eq:28}
\end{equation}
On the other hand, since each edge on the longest path in $\G_{\gen'-1}$, which determines its
diameter $L_{\gen'-1}$, is replaced in $\G_{\gen'}$ with the generator $\Ggen$ whose root nodes are
separated by $\lambda$ from each other, the diameter $L_{\gen'}$ of $\G_{\gen'}$ is given by
\begin{equation}
L_{\gen'}=\lambda L_{\gen'-1} .
\label{eq:29}
\end{equation}
Equations (\ref{eq:28}) and (\ref{eq:29}) indicate that when the radius $L_{\gen'-1}$ is multiplied by
$\lambda$, the number of nodes $\tilde{\nu}_{\gen'-1}$ is multiplied by $m_{\text{gen}}/\kappa$.
This implies that the local fractal dimension around the node $i$ is $\log
(m_{\text{gen}}/\kappa)/\log\lambda$ which is equal to $\dfmin$ of a network formed by the
single-generator model. This argument holds even if we take the limit of $\gen\to\infty$ while
keeping $\gen_{0}$ finite. In this limit, $k_{\gen}$ in Eq.~(\ref{eq:25}) diverges. Therefore, in
the thermodynamic limit where bifractality in the strict sense is established, fractality around a
hub node with an infinitely high degree is described by the local fractal dimension $\dfmin$.

Even in the thermodynamic limit, the fraction of infinite-degree hub nodes is infinitesimal, and
almost all of the nodes have finite degrees \cite{Rozenfeld07}. A substructure around a
finite-degree node located at finite distance $l_{0}$ from an infinite-degree hub node will have
the same fractality as that around the infinite-degree hub node at the scale of $l\gg l_{0}$. Thus,
the local fractal dimension around such a finite-degree node is also $\dfmin$. The fraction of
these finite-degree nodes near hubs is, however, still zero, and the overwhelming majority are the
remaining finite-degree nodes that are infinitely far from infinite-degree hub nodes. This can be
also understood from the behavior of the singularity spectrum $f(\alpha)$ which is the Legendre
transform of $\tau(q)$. Performing the Legendre transformation on $\tau(q)$ given by
Eq.~(\ref{eq:3}), we have $f(\alpha_{\text{min}})=0$ and $f(\alpha_{\text{max}})=\Df$. This implies
that the hub nodes corresponding to $\alpha_{\text{min}}(=\dfmin)$ have a point-like measure, while
the non-hub regions corresponding to $\alpha_{\text{max}}(=\dfmax)$ are extended with the same
fractal dimension as $\Df$ for the entire network. The argument on the fractality around hub nodes
cannot be applied to the substructure around such a finite-degree node $i$. This is because
$\gen-\gen_{0}$ must be finite for the degree $k_{\gen}$ of the node $i$ to be finite, and
Eq.~(\ref{eq:26}) requiring the condition $\gen'\le\gen-\gen_{0}$ is valid only in a narrow range
of $L_{\gen'}$. Considering that finite-degree nodes infinitely far from infinite-degree nodes
dominate the whole network, the local fractality around such a finite-degree node must be described
by the global fractal dimension $\Df$ which is identical to $\dfmax$. Therefore, the lower local
fractal dimension $\dfmin$ reflects the fractality of substructures around infinitely-high degree
hub nodes and finite-degree nodes at finite distances from these hub nodes, while the higher local
fractal dimension $\dfmax$ characterizes the local fractality around finite-degree nodes infinitely
far from infinite-degree hub nodes.

\begin{figure}[tttt]
    \begin{center}
    \includegraphics[width=0.9\linewidth]{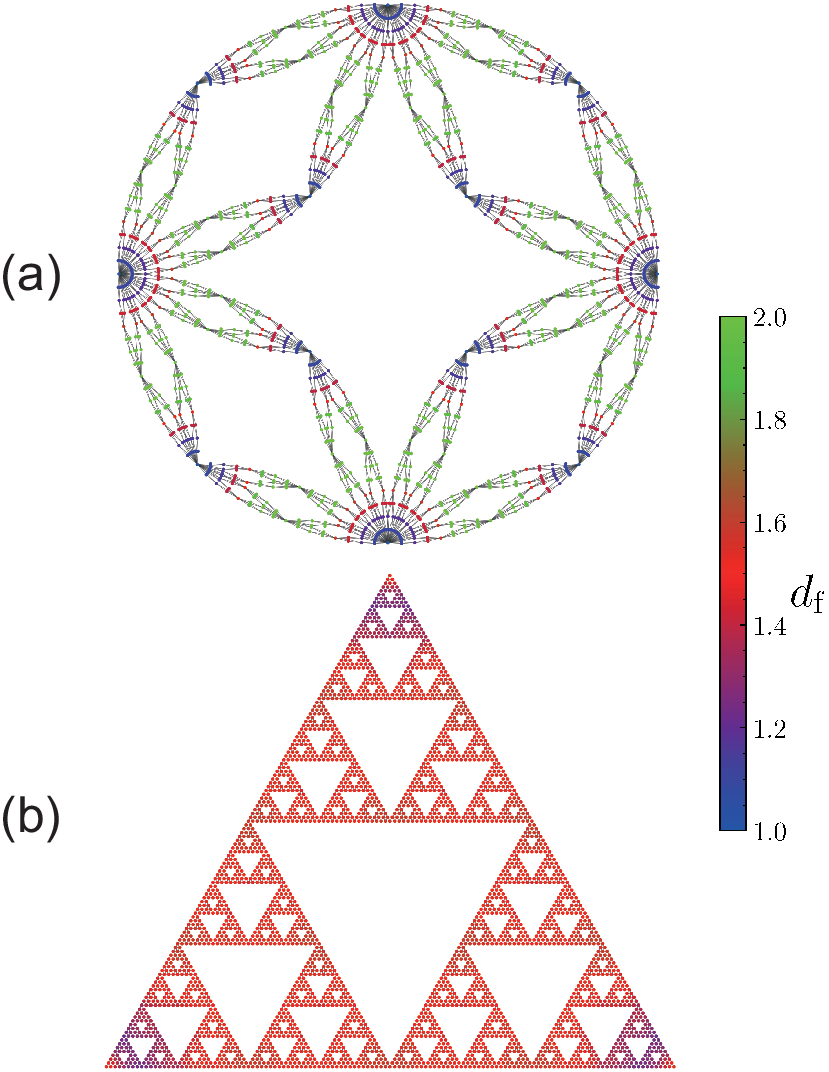}
    \caption{
Distribution of the local fractal dimension $\df$ on (a) the $6$-th generation $(2,2)$-flower
and (b) $7$-th generation Sierpinski gasket. Colors on nodes indicate values of $\df$.
}
    \label{fig:5}
    \end{center}
\end{figure}
We numerically investigate the local fractal dimension for the $6$-th generation $(2,2)$-flower
($N_{6}=2732$) to confirm the above argument. For this purpose, we count the number of nodes
$\tilde{\nu}_{i}(l)$ within distance $l$ from every node $i$ in the $(2,2)$-flower and
calculate the local fractal dimension $\df(i)$ according to the relation
$\tilde{\nu}_{i}(l)\propto l^{\df(i)}$\cite{Xiao21}. Figure \ref{fig:5}(a) indicates the value of
$\df(i)$ at each node $i$ by color. As expected, local fractal dimensions $\df(i)$ take values
close to $\dfmin=1$ on high-degree nodes and close to $\dfmax=2$ on low-degree nodes away from the
high-degree nodes. It is due to the finite-size effect that $\df(i)$'s on some nodes take
intermediate values between $\dfmin$ and $\dfmax$. Regions around these nodes do not actually have
fractality described by the intermediate $\df$, but the least-squares fit for $\tilde{\nu}_{i}(l)$
of the finite-size network merely evaluates the dimensional crossover from $\dfmax$ to $\dfmin$ as
if having the intermediate dimension. The scale-free property is crucial for the distribution of
the local fractal dimension. In a fractal network without the scale-free property, the local
fractality around any node is described by the global fractal dimension, as shown by
Fig.~\ref{fig:5}(b). In this figure, $\df(i)$ on every node in the $7$-th generation Sierpinski
gasket ($N_{7}=3282$) is indicated. The computed $\df(i)$'s take values close to the global fractal
dimension $\Df=\log 3/\log 2(=1.585)$ regardless of the node, except for nodes near the three
corners affected by the finite-size effect.

\section{Conclusion}
\label{sec:conclusion}

In this work we have studied the structural multifractality of extensive classes of fractal
scale-free networks (FSFNs) and conjectured that any FSFN possesses a bifractal structure
characterized by two local fractal dimensions. It has been reported by the previous work
\cite{Furuya11} that if the number of nodes in each of boxes covering an FSFN minimally is
proportional to the number of adjacent boxes, namely Eq.~(\ref{eq:2}) holds, the network is
bifractal. Based on this proposition, we first showed that deterministic hierarchical FSFNs formed
by the single-generator model exhibit the bifractal nature in their structures. This result was
confirmed for several deterministic hierarchical FSFNs by comparing the mass exponents $\tau(q)$
predicted theoretically with those calculated numerically. Next, we examined stochastic
hierarchical FSFNs which have hierarchical structures but are not deterministic. In order to
construct such FSFNs, we proposed the multi-generator model in which every edge in the previous
generation network is recursively replaced with one of multiple generators with a certain
probability. Our analytical argument reveals that FSFNs formed by this model also have bifractal
structures. The bifractal profile of $\tau(q)$ for these networks has been verified numerically by
employing several combinations of multiple generators. Furthermore, we showed that non-hierarchical
FSFNs exhibit bifractality, by investigating the giant component of a scale-free random network at
the percolation critical point. By evaluating the conditional probability describing the long-range
degree correlation of the giant component and by utilizing it to count the number of nodes in a box
and the number of neighboring boxes, we verified Eq.~(\ref{eq:2}). Finally, we demonstrated that
the mass exponents for two real-world FSFNs are consistent with the bifractal profiles of $\tau(q)$
predicted by their degree exponents and global fractal dimensions. From these results, we
conjecture that any FSFN is bifractal. Moreover, we studied which substructures of an FSFN are
characterized by its local fractal dimensions $\dfmin$ and $\dfmax$. Our arguments clarified that
$\dfmin$ describes substructures around infinitely high-degree hub nodes and finite-degree nodes at
finite distances from these hub nodes, while $\dfmax$ describes the local fractality around
finite-degree nodes infinitely far from infinite-degree hub nodes.

The bifractality in the strict sense is realized only in infinitely large FSFNs. However, even in a
finite-size FSFN, the fractality cannot be described only by the global fractal dimension $\Df$.
Local structures near high-degree hub nodes are characterized by $\dfmin$ and the fractality near
low-degree nodes far from hub nodes are quantified by $\dfmax$. Meanwhile, we expect a dimensional
crossover in the vicinity of a low-degree node not far from the hub or a node with an intermediate
degree. Therefore, the multifractal analysis for a finite-size FSFN makes it appear as if the
network possesses multifractality described by an infinite number of local fractal dimensions, but
this is merely a finite-size effect of the bifractal nature.

It is important to study how the structural bifractality affects phenomena occurring on FSFNs.
Random walks on FSFNs, for example, can be influenced by bifractality of the networks.
In fact, it has been reported that there are two distinct values of the spectral
dimension $d_{\text{s}}$ characterizing the time dependence of the return-to-origin probability of
a random walker \cite{Hwang12a,Hwang12b,Hwang13}, though the distance dependence of the
first-passage time is described only by a single walk dimension $d_{\text{w}}$ \cite{Rozenfeld07}.
Future research should clarify the relevance of these two spectral dimensions and the bifractal
nature of the network, which will lead to a systematic understanding of various dynamics on
bifractal networks.

\begin{acknowledgements}
The authors thank S.~Mizutaka for fruitful discussions. J.Y. also acknowledges insightful comments
from T.~Hasegawa. This work was supported by a Grant-in-Aid for Scientific Research (No.~22K03463)
from the Japan Society for the Promotion of Science. J.Y. was financially supported by the Itoh
Foundation for International Education Exchange.
\end{acknowledgements}

\appendix
\section{Derivation of Eq.~(\ref{eq:3})}
\label{appendix:a}

Here we derive the bifractal mass exponent $\tau(q)$, namely Eq.~(\ref{eq:3}), under the assumption
Eq.~(\ref{eq:2}). Given an FSFN $\G$ of $N$ nodes, let us cover $\G$ with the minimum number
$N_{\text{B}}(l)$ of boxes of diameter $l$, and assume that Eq.~(\ref{eq:2}) holds for each box
$b$. Equation (\ref{eq:2}) gives the relation,
\begin{equation}
\nu_{b}=\frac{\langle \nu_{b}\rangle}{\langle k_{b}\rangle}k_{b} ,
\label{eq:a1}
\end{equation}
where $\langle \nu_{b}\rangle$ and $\langle k_{b}\rangle$ are the mean values of $\nu_{b}$ and
$k_{b}$ averaged over the boxes, respectively. Since $\langle \nu_{b}\rangle=N/N_{\text{B}}(l)$, we
have $\nu_{b}=k_{b}N/\langle k_{b}\rangle N_{\text{B}}(l)$ and the box measure $\mu_{b}=\nu_{b}/N$
is given by
\begin{equation}
\mu_{b}=\frac{k_{b}}{\langle k_{b}\rangle N_{\text{B}}(l)} .
\label{eq:a2}
\end{equation}
The number of boxes $N_{\text{B}}(l)$ is proportional to $l^{-\Df}$. Then, the box measure
$\mu_{b}$ is written as
\begin{equation}
\mu_{b}(k_{b})\propto \frac{k_{b}}{\langle k_{b}\rangle}l^{\Df} .
\label{eq:a3}
\end{equation}
Therefore, the $q$-th moment $Z_{q}(l)$ defined by Eq.~(\ref{eq:1}) is calculated as
\begin{eqnarray}
Z_{q}(l)=\sum_{b}\mu_{b}^{q}(k_{b}) &\simeq& N_{\text{B}}(l) \int_{0}^{\infty}\mu_{b}^{q}(k_{b})P(k_{b}) \,dk_{b} \nonumber \\
&\propto& \frac{l^{(q-1)\Df}}{\langle k\rangle^{q}} \int_{0}^{\infty} k^{q}P(k)\,dk ,
\label{eq:a4}
\end{eqnarray}
where $P(k_{b})$ is the degree distribution of the super-node $b$ in the renormalized network of
$\G$, which is the same as the degree distribution $P(k)\propto k^{-\gamma}$ of the original
network $\G$ due to the fractal property of $\G$.

In the case of $q-\gamma<-1$, the integral in Eq.~(\ref{eq:a4}), namely the $q$-th moment of $k$,
converges to a constant value. Thus, we have $Z_{q}(l)\propto l^{(q-1)\Df}$, which implies that
$\tau(q)=(q-1)\Df$ for $q<\gamma-1$. For $q\ge \gamma-1$, however, the integral in
Eq.~(\ref{eq:a4}) must be evaluated by considering that $N_{\text{B}}(l)$ is sufficiently large but
finite. In this case, the degree of the renormalized network is bounded by the natural cut-off
$k_{\text{c}}(l)\propto [N_{\text{B}}(l)]^{1/(\gamma-1)}\propto l^{-\Df/(\gamma-1)}$
\cite{Dorogovtsev02}. Hence, the $q$-th moment becomes
\begin{eqnarray}
Z_{q}(l)&\propto& \frac{l^{(q-1)\Df}}{\langle k\rangle^{q}} \int^{k_{\text{c}}(l)} k^{q-\gamma}\,dk \nonumber \\
&\propto& l^{\Df q(\gamma-2)/(\gamma-1)} .
\label{eq:a5}
\end{eqnarray}
This gives $\tau(q)=q\Df (\gamma-2)/(\gamma-1)$ for $q\ge \gamma-1$.

\section{Properties of networks formed by the multi-generator model}
\label{appendix:b}

Let us consider properties of the $\gen$-th generation network $\G_{\gen}$ formed by the
multi-generator model. If $\G_{\gen-1}$ contains $M_{\gen-1}$ edges, $p^{(i)}M_{\gen-1}$ edges, on
average, in $\G_{\gen-1}$ are replaced with $\Ggen^{(i)}$ in the growth process from $\G_{\gen-1}$
to $\G_{\gen}$, and these edges proliferate to $p^{(i)}M_{\gen-1}m_{\text{gen}}^{(i)}$ in
$\G_{\gen}$, where $m_{\text{gen}}^{(i)}$ is the number of edges in $\Ggen^{(i)}$. Thus, the number
of edges in $\G_{\gen}$ is given by $M_{\gen}=M_{\gen-1}\mgenav$, where
$\mgenav=\sum_{i}^{s}m_{\text{gen}}^{(i)}p^{(i)}$ is the mean number of edges averaged over the
multiple
generators. Solving this recurrence equation under the condition $M_{0}=1$, the number of edges
in $\G_{\gen}$ is given by
\begin{equation}
M_{\gen}=\mgenav^{\gen} .
\label{eq:b1}
\end{equation}
Regarding the number of nodes, $p^{(i)}M_{\gen-1}n_{\text{rem}}^{(i)}$ nodes in $\G_{\gen}$ emerges
from $p^{(i)}M_{\gen-1}$ edges in $\G_{\gen-1}$ by the edge replacement operation with
$\Ggen^{(i)}$, where $n_{\text{rem}}^{(i)}$ is the number of remaining nodes of $\Ggen^{(i)}$. The
total number of newly emerged nodes in $\G_{\gen}$ is then given by $M_{\gen-1}\nremav$, where
$\nremav=\sum_{i}^{s}n_{\text{rem}}^{(i)}p^{(i)}$ is the mean number of remaining nodes averaged
over the multiple generators. Thus, using Eq.~(\ref{eq:b1}), the number of nodes $N_{\gen}$ in
$\G_{\gen}$ is written as
\begin{equation}
N_{\gen}=N_{\gen-1}+\nremav \mgenav^{\gen-1} .
\label{eq:b2}
\end{equation}
The solution of this recurrence equation under the initial condition $N_{0}=2$ is given as
\begin{equation}
N_{\gen}=2+\frac{\nremav\left(\mgenav^{\gen}-1\right)}{\mgenav-1} .
\label{eq:b3}
\end{equation}
This expression is the same as Eq.~(\ref{eq:5}) if we replace $\nremav$ and $\mgenav$ by
$n_{\text{rem}}$ and $m_{\text{gen}}$, respectively. Since $N_{\gen}$ is proportional to
$\mgenav^{\gen}$ for $\gen\gg 1$ as can be seen from Eq.~(\ref{eq:b3}), we have
\begin{equation}
N_{\gen}=\mgenav N_{\gen-1},
\label{eq:b4}
\end{equation}
for high generation networks.

Next, we consider the asymptotic form of the degree distribution $P(k)$ for sufficiently large $k$
and $\gen\gg 1$. Since $p^{(i)}k$ edges out of the $k$ edges incident to a node of degree $k$ are
replaced with $\Ggen^{(i)}$, these $p^{(i)}k$ edges increase the degree of the node by
$\kappa^{(i)}p^{(i)}k$ in the next generation, where $\kappa^{(i)}$ is the degree of the root node
in $\Ggen^{(i)}$. Therefore, the number of nodes, $N_{\gen-1}(k)$, of degree $k$ in $\G_{\gen-1}$
is the same as that of nodes of degree $\sum_{i}\kappa^{(i)}p^{(i)}k$ in $\G_{\gen}$, namely
$N_{\gen-1}(k)=N_{\gen}(\kappaav k)$, where $\kappaav=\sum_{i}\kappa^{(i)}p^{(i)}$ is the mean
degree of the root nodes averaged over the generators. Using the asymptotic degree distribution
$P(k)$ for $\gen\gg 1$, this relation can be written as $N_{\gen-1}P(k)=\kappaav N_{\gen}P(\kappaav k)$,
and it follows from Eq.~(\ref{eq:b4}) that
\begin{equation}
P(k)=\mgenav\kappaav P(\kappaav k) .
\label{eq:b5}
\end{equation}
The solution of this functional equation is $P(k)\propto k^{-\gamma}$ with
\begin{equation}
\gamma=1+\frac{\log \mgenav}{\log \kappaav} .
\label{eq:b6}
\end{equation}
Thus, networks formed by this stochastic model possess the scale-free property.

We can also show that $\G_{\gen}$ with $\gen\gg 1$ has a fractal structure. In order to examine the
fractal property of $\G_{\gen}$, we consider the distance $l_{\gen}$ between two specific nodes in
the network $\G_{\gen}$. Considering that edge replacements are performed randomly, we can expect a
relation of $l_{\gen}=\lambdaav l_{\gen-1}$ with some coefficient $\lambdaav$ for $\gen\gg 1$,
where $l_{\gen-1}$ is the distance between the same node pair in $\G_{\gen-1}$. It should be
emphasized that the mean inter-root-node distance $\lambdaav$ is not given by the simple average
$\sum_{i}\lambda^{(i)}p^{(i)}$, where $\lambda^{(i)}$ is the distance between root nodes in the
generator $\Ggen^{(i)}$, because the shortest path itself may change due to the edge replacements.
The correct $\lambdaav$ can be calculated using the concept of renormalization. Since $\lambdaav$
is independent of the choice of node pairs, let $l_{\gen}$ be the distance between two nodes in
$\G_{\gen}$ corresponding to the two terminal nodes of the zeroth-generation graph consisting of
two nodes and a single edge. From a renormalization perspective, $\G_{\gen}$ takes a structure in
which each edge of the generators is replaced with one of the possible structures from
$\G_{\gen-1}$ with equal probability. Therefore, we have
\begin{equation}
l_{\gen}=\sum_{i=1}^{s}p^{(i)}l_{\gen}^{(i)} ,
\label{eq:b7}
\end{equation}
with
\begin{equation}
l_{\gen}^{(i)}=F^{(i)}\left(\{l_{\gen-1}^{(j)}\},\{p^{(j)}\} \right) .
\label{eq:b8}
\end{equation}
Here, $l_{\gen}^{(i)}$ is the average distance between root nodes in the generator $\Ggen^{(i)}$
in which each edge has one of the lengths $\{l_{\gen-1}^{(j)}:1\le j\le s\}$ with probability
$p^{(j)}$, and $F^{(i)}$ is the function mapping from $\{l_{\gen-1}^{(j)}\}$ to $l_{\gen}^{(i)}$.
Given the structures of the generators, the functional forms of $\{F^{(i)}\}$ are determined, and
we can iteratively compute $l_{\gen}^{(i)}$ by Eq.~(\ref{eq:b8}) and then obtain $l_{\gen}$ from
Eq.~(\ref{eq:b7}). From the relation $l_{\gen}=\lambdaav l_{\gen-1}$, the mean inter-root-node
distance $\lambdaav$ is given by
\begin{equation}
\lambdaav=\lim_{\gen\to\infty}\frac{l_{\gen}}{l_{\gen-1}} .
\label{eq:b9}
\end{equation}
Recalling that $\lambdaav$ does not depend on the choice of node pairs, for the diameter $L_{\gen}$
of the network $\G_{\gen}$ we have
\begin{equation}
L_{\gen}=\lambdaav L_{\gen-1} .
\label{eq:b10}
\end{equation}
for $\gen\gg 1$. Although the two nodes determining the diameter of $\G_{\gen}$ may be slightly
distant from the nodes giving the diameter of $\G_{\gen-1}$, Eq.~(\ref{eq:b10}) is valid for
sufficiently large $\gen$.
The combination of Eqs.~(\ref{eq:b4}) and (\ref{eq:b10}) implies that as
the network diameter becomes $\lambdaav$ times larger, the number of nodes increases by a factor of
$\mgenav$. This leads to the relation $N_{\gen}\propto L_{\gen}^{\Df}$ with
\begin{equation}
\Df=\frac{\log \mgenav}{\log \lambdaav} .
\label{eq:b11}
\end{equation}

As an example, let us calculate $\Df$ for the combination of the black generators shown in
Fig.~\ref{fig:3}. Setting $\Ggen^{(1)}$ and $\Ggen^{(2)}$ as the generators with $4$ and $6$ nodes,
respectively, the functions $F^{(1)}$ and $F^{(1)}$ are given by
\begin{align}
F^{(1)}&(l_{\gen}^{(1)},l_{\gen}^{(2)},p)=                                  \nonumber \\
 &2 l_{\gen}^{(1)}                                        p^{4}
 +4 \min(l_{\gen}^{(1)}+l_{\gen}^{(2)},2l_{\gen}^{(1)})   p^{3}(1-p)        \nonumber \\
+&2 \min(2l_{\gen}^{(1)},2l_{\gen}^{(2)})                 p^{2}(1-p)^{2}
 +4 (l_{\gen}^{(1)}+l_{\gen}^{(2)})                       p^{2}(1-p)^{2}    \nonumber \\
+&4 \min(l_{\gen}^{(1)}+l_{\gen}^{(2)},2l_{\gen}^{(2)})   p(1-p)^{3}
 +2 l_{\gen}^{(2)}(1-p)^{4} ,
\label{eq:b12} \\
F^{(2)}&(l_{\gen}^{(1)},l_{\gen}^{(2)},p)=                 \nonumber \\
&  3l_{\gen}^{(1)}                       p^{5}
 + 2\cdot 3l_{\gen}^{(1)}                p^{4}(1-p)
 + 3(2l_{\gen}^{(1)}+l_{\gen}^{(2)})     p^{4}(1-p)        \nonumber \\
+& 6(2l_{\gen}^{(1)}+l_{\gen}^{(2)})     p^{3}(1-p)^{2}
 + 3l_{\gen}^{(1)}                       p^{3}(1-p)^{2}    \nonumber \\
+& 3(l_{\gen}^{(1)}+2l_{\gen}^{(2)})     p^{3}(1-p)^{2}
+ 3l_{\gen}^{(2)}                        p^{2}(1-p)^{3}    \nonumber \\
+& 6(l_{\gen}^{(1)}+2l_{\gen}^{(2)})     p^{2}(1-p)^{3}
 + 3(2l_{\gen}^{(1)}+l_{\gen}^{(2)})     p^{2}(1-p)^{3}    \nonumber \\
+& 2\cdot 3l_{\gen}^{(2)}                p(1-p)^{4}
 + 3(l_{\gen}^{(1)}+2l_{\gen}^{(2)})     p(1-p)^{4}        \nonumber \\
+& 3l_{\gen}^{(2)}                       (1-p)^{5} ,
\label{eq:b13}
\end{align}
where $p$ is the edge replacement probability for $\Ggen^{(1)}$, and the function $\min(x,y)$
returns the minimum value of $x$ and $y$. These functions $F^{(1)}$ and $F^{(2)}$enable us to
calculate the mean distance $\lambdaav$ from Eqs.~(\ref{eq:b7})-(\ref{eq:b9}). For $p=1/2$ as in
the case of Fig.~\ref{fig:3}, for example, $\lambdaav$ given by Eq.~(\ref{eq:b9}) rapidly converges
to $\lambdaav=2.408$. Since $\mgenav=9/2$ for these $\Ggen^{(1)}$ and $\Ggen^{(2)}$, the fractal
dimension for $p=1/2$ is $\Df=1.712$. This value, of course, varies with $p$ from
$\Df=\log 5/\log 2=1.465$ ($p=0$) to $\Df=2$ ($p=1$), which implies that we can continuously
control the fractality of FSFNs by using the multi-generator model.

The above results show that networks built by the multi-generator model are FSFNs as in the case of
the single-generator model. We should note that the expressions of $M_{\gen}$, $N_{\gen}$,
$\gamma$, and $\Df$ are the same as those for the single-generator model if we replace quantities
characterizing the single generator by their mean values for the multiple generators.


\end{document}